\documentclass[sigconf]{acmart}

\usepackage{amsmath}
\usepackage{booktabs}
\usepackage{graphicx}
\usepackage{xcolor}
\usepackage{listings}
\usepackage{algorithm}
\usepackage{algorithmic}
\usepackage{subcaption}
\usepackage{multirow}
\usepackage{tikz}
\usetikzlibrary{shapes,arrows,positioning,fit,calc}

\newcommand{\cosine}{\mathrm{cos\text{-}sim}}
\newcommand{\textemb}{\textsf{TextEmb}}
\newcommand{\idem}{\textsf{IdEmb}}

\AtBeginDocument{}
\setcopyright{none}                % suppress the default copyright statement
\renewcommand\footnotetextcopyrightpermission[1]{} % drop the copyright/permission footnote

\acmConference[RecSys '26]{Proceedings of the 20th ACM Conference
on Recommender Systems}{September 28--October 2, 2026}
{Minneapolis, MN, USA}
\copyrightyear{2026}
\acmYear{2026}
\begin{document}

\title[Personalizing Incremental Video Search]{Personalizing Incremental Video Search with Hybrid Text and ID Embeddings}

\titlenote{Accepted to the Industry Track of the 20th ACM Conference on Recommender Systems (RecSys 2026). This is the authors' accepted manuscript; the final version will appear in the ACM Digital Library.}

\author{Vivek Kanojiya}
\affiliation{%
  \institution{Apple}
  \city{Seattle}
  \state{WA}
  \country{USA}}
\email{vkanojiya@apple.com}

\author{Vishalaksh Aggarwal}
\affiliation{%
  \institution{Apple}
  \city{Seattle}
  \state{WA}
  \country{USA}}
\email{vishalaksh@apple.com}

\author{Daeho Baek}
\affiliation{%
  \institution{Apple}
  \city{Cupertino}
  \state{CA}
  \country{USA}}
\email{daeho@apple.com}

\author{Lyndon Kennedy}
\affiliation{%
  \institution{Apple}
  \city{Cupertino}
  \state{CA}
  \country{USA}}
\email{lyndon_kennedy@apple.com}

\author{Xuetao Yin}
\affiliation{%
  \institution{Apple}
  \city{Cupertino}
  \state{CA}
  \country{USA}}
\email{xuetao.yin@apple.com}

%% ------------------------------------------------------------------
\begin{abstract}
Incremental video search requires high-quality ranking after each keystroke, where intent is often underspecified (e.g., 1--3 character prefixes). We present a personalization system for Apple TV search that combines complementary semantic and collaborative signals at ranking time. Our approach learns two item embedding spaces: (i) a text-based multilingual encoder (\textemb{}) fine-tuned on co-engagement triplets via contrastive learning, and (ii) an ID-based collaborative embedding model (\idem{}) trained on interaction-derived positives. At serving time, we construct user representations from recent watch history and inject text- and ID-based user--item cosine similarities into a pairwise XGBoost ranker.

We evaluate the system with temporally held-out offline datasets and a three-week online controlled experiment. Offline, for sessions with user history, the personalized ranker improves NDCG@10 by 2.99\% and MRR by 3.30\% over the non-personalized baseline. Crucially, slice analyses show that personalization is most needed in \emph{incremental search}, where intent is still forming: on ambiguous prefix queries (1--3 characters), NDCG@10 lift is +8.63\%, versus only +1.46\% on longer, more fully specified queries. Users with longer watch histories benefit more from personalization than newer users: NDCG lift rises from +2.13\% for users with 1--5 history items to +4.37\% for users with 51--100. This larger lift occurs even though baseline relevance is lower for long-history cohorts (NDCG@10 drops from 0.733 to 0.680), indicating that personalization adds the most value where default ranking underperforms. Online, treatment yields statistically significant gains of +1.14\% tap-through rate and +1.23\% conversion rate, with a 2.91\% improvement in converted-item rank position. We further analyze coverage--precision trade-offs between semantic and collaborative embeddings through ablations isolating each signal, and evaluate embedding quality on a held-out corpus with LLM-judged similarity labels to reduce click/exposure bias.
\end{abstract}



\maketitle

%% ==================================================================
\section{Introduction}
\label{sec:intro}
%% ==================================================================

Search in large-scale video catalog applications differs from classical web search in a way that makes personalization especially important: ranking is \textit{incremental}. The system must return useful results after every keystroke, including short prefixes where lexical evidence is weak but user intent already matters. In this regime, the first 1--3 characters create a \textit{signal vacuum}: many plausible candidates share nearly indistinguishable text evidence, so non-personalized ranking tends to fall back to global popularity and broad behavioral priors rather than the current user's taste.

This short-prefix regime is the core problem addressed in this paper. Our central empirical finding is that personalization matters far more on these ambiguous incremental-search inputs than on longer, more fully specified queries: offline, the proposed personalized ranker improves NDCG@10 by \textbf{+8.63\%} on prefix queries of length 1--3, compared with only \textbf{+1.46\%} on longer queries. We therefore frame the problem not simply as ``building a better personalized ranker,'' but as identifying the right personalization signals for the low-lexical-information regime of incremental search.

\begin{figure*}[t]
\centering
\includegraphics[width=\textwidth,height=0.42\textheight,keepaspectratio]{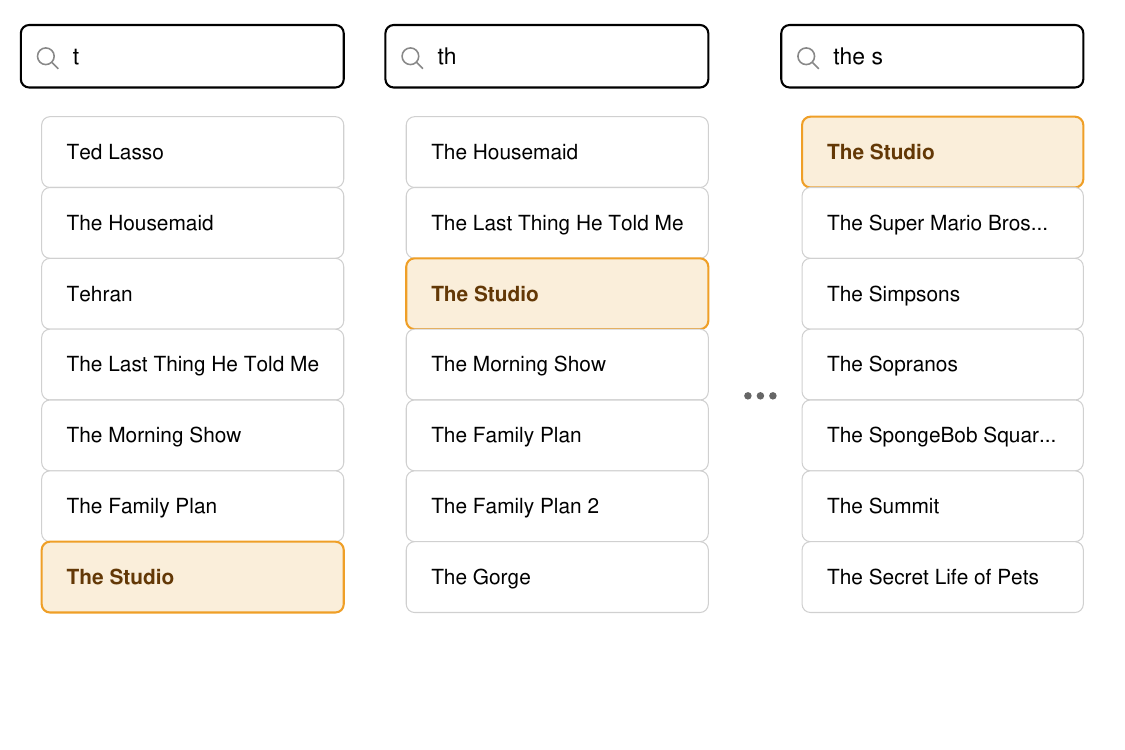}
\caption{UI-style illustration of non-personalized prefix-search ambiguity. For short prefixes such as \texttt{t} or \texttt{th}, many candidate titles may match, so personalization is needed to rank the intended item earlier instead of requiring users to type additional characters before it becomes easy to find.}
\Description{Autocomplete UI showing non-personalized search states for the prefixes t, th, and the s. The intended title, The Studio, appears low in the ranking for short prefixes and becomes easier to find only after more characters are typed.}
\label{fig:incremental_ui}
\end{figure*}

Figure~\ref{fig:incremental_ui} illustrates non-personalized incremental search behavior using three prefixes of increasing length from the same query (\texttt{t} $\rightarrow$ \texttt{th} $\rightarrow$ \texttt{the s}). In a non-personalized search experience, users may need to type several additional characters before the intended item becomes easy to find.

We compare against a non-personalized baseline ranker driven mainly by aggregated behavioral and lexical signals, so users in the same geography would see a shared default ranking for a query. While this baseline is robust for fully specified navigational queries, it underperforms precisely where incremental search is hardest: early prefixes with many plausible matches and insufficient lexical separation.

Our thesis is that \textit{hybrid embeddings are the right answer for this regime}. Text-based embeddings provide semantic coverage and cold-start robustness when interaction data are sparse, while ID-based embeddings contribute high-precision collaborative structure when co-engagement signals are dense. Injecting both signals into a ranker lets the system personalize exactly where lexical matching is least informative, without sacrificing coverage or latency.

We target five constraints that make this problem challenging in practice:
\begin{itemize}
  \item \textbf{Ambiguity at short prefixes.} Candidate sets are large while lexical discriminative power is low.
  \item \textbf{Broad-intent queries with many valid matches.} Generic queries (e.g., ``comedy'') can return a large set of plausible candidates, making personalization critical for surfacing the most relevant results for each user.
  \item \textbf{Coverage across the catalog.} Personalization should work for both popular and long-tail content.
  \item \textbf{Cold-start resilience.} Newly launched items lack interaction history and cannot rely on collaborative signals alone.
  \item \textbf{Strict serving budgets.} Personalization must satisfy low-latency constraints at scale.
\end{itemize}

To address these constraints, we introduce a hybrid personalization stack that combines semantic and collaborative embeddings within a learning-to-rank pipeline. Our core contributions are
\begin{enumerate}
  \item \textbf{Problem framing for incremental-search personalization.} We show that personalization has disproportionate value on short-prefix queries, with NDCG@10 improving by +8.63\% on 1--3 character queries versus +1.46\% on longer queries.
  \item \textbf{Hybrid embedding formulation for the short-prefix regime.} We learn two complementary item spaces: a text-semantic encoder (\textemb{}) and an ID-collaborative model (\idem{}), and show through ablations that each contributes non-redundant ranking signal.
  \item \textbf{Online user representation for low-latency ranking.} We compute user embeddings from previously watched items with bounded-memory aggregation, and expose per-candidate text/ID cosine features to the ranker with minimal latency overhead.
  \item \textbf{Ranker integration.} We inject hybrid personalization signals into a pairwise XGBoost model trained on large scale user engagement dataset, preserving compatibility with an existing multi-stage retrieval and ranking stack.
  \item \textbf{Evaluation across offline and online regimes.} We report temporal hold-out metrics (NDCG@10, MRR@10, converted-item position) and a three-week online A/B test demonstrating statistically significant engagement gains.
  \item \textbf{Embedding-quality analysis with reduced click bias.} We construct a held-out evaluation corpus via a large language model (LLM), decoupling embedding quality assessment from exposure bias in session-derived data.
\end{enumerate}

Table~\ref{tab:headline_results} summarizes the headline results. The prefix-query lift is the central result: it shows that incremental search is the regime where personalization contributes the most value.

\begin{table}[t]
\centering
\caption{Headline results. Prefix-query lift is the strongest evidence that personalization is most valuable in the low-lexical-information regime of incremental search.}
\label{tab:headline_results}
\begin{tabular}{p{0.44\columnwidth}p{0.22\columnwidth}p{0.18\columnwidth}}
\toprule
\textbf{Setting} & \textbf{Metric} & \textbf{Lift} \\
\midrule
Offline, all sessions with history & NDCG@10 & +2.99\% \\
Offline, prefix queries (1--3 chars) & NDCG@10 & +8.63\% \\
Offline, longer queries (>3 chars) & NDCG@10 & +1.46\% \\
Online A/B test & Tap-through rate & +1.14\% \\
Online A/B test & Conversion rate & +1.23\% \\
\bottomrule
\end{tabular}
\end{table}

%% ==================================================================
\section{Related Work}
\label{sec:related}
%% ==================================================================

\subsection{Learning to Rank}
Learning-to-rank methods for search include pointwise, pairwise, and listwise formulations~\cite{liu2009learning}. Pairwise approaches such as RankNet/LambdaRank remain strong practical choices in industrial systems~\cite{burges2006learning,burges2010ranknet}. In high-throughput applications, tree-based models with pairwise objectives (e.g., LambdaMART, XGBoost with \texttt{rank:pairwise}) provide an effective trade-off between ranking quality, feature flexibility, and serving efficiency~\cite{chen2016xgboost}. Our work builds on this line by adding hybrid embedding signals to an existing pairwise ranker.

\subsection{Personalized Search}
Personalized web search has a long history~\cite{dou2007large,teevan2005personalizing}. Early work used long-term user interest profiles derived from query logs; later work distinguished short-term and long-term behavioral signals~\cite{bennett2012modeling}. Recent industrial evidence from streaming search at Netflix further highlights the practical impact of personalization in large-scale media systems~\cite{ostuni2023search}. A key challenge in this line of work---also central to our setting---is balancing personalization strength against degradation for navigational queries where all users share the same intent. Our system explicitly addresses this by treating personalization signals as additive features within an existing relevance-first ranker.

\subsection{Embedding-based Retrieval and Dense Bi-Encoders}
Dual-encoder architectures such as DSSM~\cite{huang2013learning} established independent query/document encoding for scalable relevance modeling. Later two-tower systems improved industrial retrieval through large-batch training and ANN serving~\cite{yi2019sampling}. Dense retrieval methods (DPR~\cite{karpukhin2020dense} and large dual encoders~\cite{ni2021large}) demonstrated that bi-encoder models pre-trained on natural language can generalize well to new domains via contrastive fine-tuning. Sentence-BERT~\cite{reimers2019sentencebert} extended this to sentence-level semantic similarity; we adapt this line of work to multilingual media metadata with a domain-specific fine-tuning objective.

\subsection{Collaborative and Hybrid Recommendation}
Collaborative methods from matrix factorization~\cite{koren2009matrix} to neural collaborative filtering~\cite{he2017neural} and item2vec-style representations~\cite{barkan2016item2vec} capture behavioral proximity effectively for head items. Hybrid recommenders combine collaborative and content-based signals to improve robustness under sparsity and cold-start~\cite{burke2002hybrid}. Our approach instantiates this hybrid principle in search ranking. Contrastive self-supervised objectives have also been applied to recommendation~\cite{yao2021self,xie2022contrastive}, motivating our use of a pairwise hinge loss for embedding training.

\subsection{Session-based and Sequential Recommendation}
Session-based models~\cite{hidasi2016session,li2017neural} capture short-term intent from recent interactions without requiring long-term user profiles. Sequential models such as SASRec~\cite{kang2018self} and BERT4Rec~\cite{sun2019bert4rec} apply self-attention over the history to model temporal dynamics. Our current system uses mean pooling over watch history, an approach that has also been used in prior large-scale industrial recommendation systems to build user embeddings by averaging watch-history item embeddings~\cite{covington2016deep}. We identify this approach of mean pooling as a limitation and target for future work; however, mean pooling provides a simple, low-latency baseline that already yields substantial gains (Section~\ref{sec:offline}).

\subsection{Personalization in Large-Scale Media Systems}
Large-scale industrial recommender stacks combine user history with multi-stage ranking and deep representation learning~\cite{gomez2015netflix,elkahky2015multi,covington2016deep}. One especially relevant prior system~\cite{covington2016deep} builds a user embedding by averaging item embeddings from watch history and uses it in a two-stage retrieve-then-rank pipeline. Recent industrial search work likewise demonstrates the value of personalization at scale~\cite{ostuni2023search}; our work differs in focusing specifically on the per-keystroke incremental regime, where lexical signal is weakest and personalization most needed. Our contribution further complements behavioral embeddings with text-semantic embeddings to handle cold-start and long-tail content.

%% ==================================================================
\section{System Overview}
\label{sec:system}
%% ==================================================================

\begin{figure}[t]
\centering
\includegraphics[width=\columnwidth]{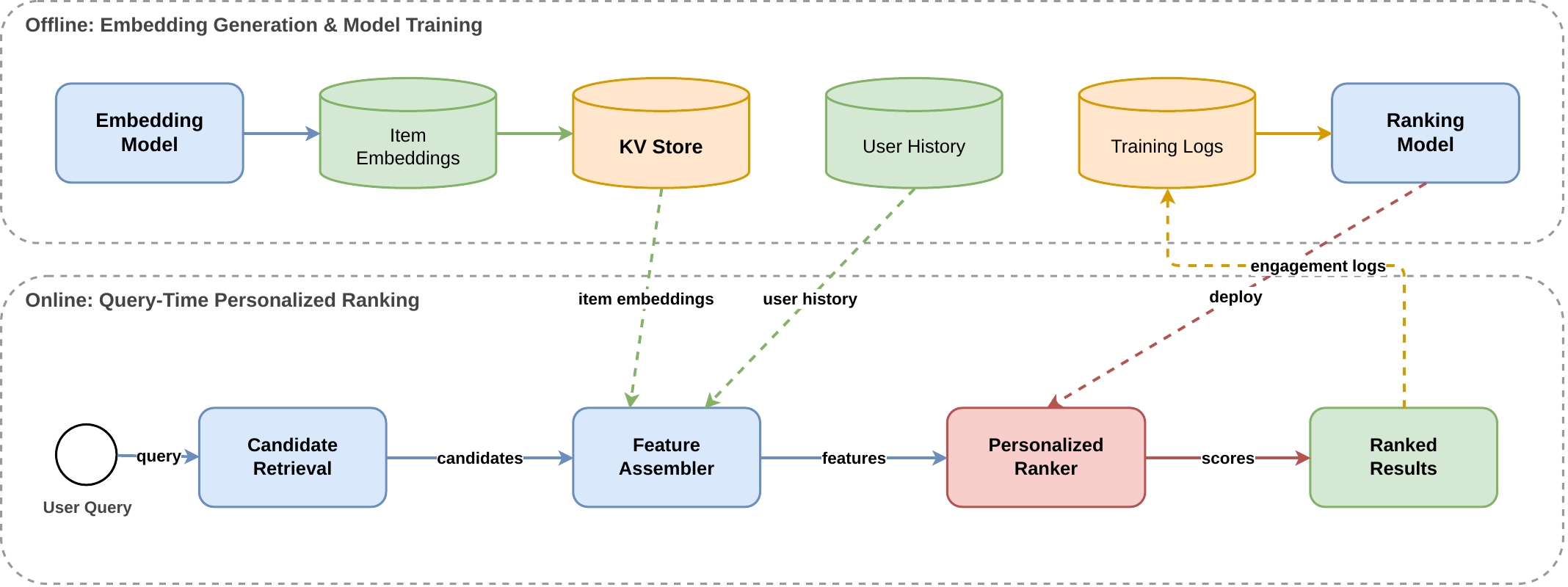}
\caption{High-level overview of the personalized search system. Item embeddings and the ranking model are produced offline, while query-time ranking combines retrieved candidates with cached embeddings to produce personalized results.}
\Description{High-level architecture diagram of personalized search showing offline embedding and ranker preparation, plus online candidate retrieval and re-ranking using cached embeddings to generate personalized results.}
\label{fig:arch}
\end{figure}

Figure~\ref{fig:arch} shows the end-to-end architecture. The system comprises three phases.

\textbf{Offline Embedding Generation.}
Item embeddings are generated offline over the full video catalog at a fixed refresh cadence using item identifiers and metadata signals. We materialize these embeddings in two serving surfaces: (i) a key--value store indexed by item ID for low-latency online lookup, and (ii) the search index document metadata so embeddings are retrieved directly during candidate recall.

\textbf{User History.}
User watch history is maintained in a low-latency key-value store, refreshed on a regular cadence from interaction logs. Per user, we retain a bounded set of the most recent play events over a long-horizon window; episode-level interactions are mapped to the parent show, and duplicate items are deduplicated to the most recent play.

\textbf{Online serving.}
At query time, the search stack retrieves user history from the key-value store and forms user representations via mean pooling over historical item embeddings. The ranker consumes both user vectors and per-candidate item embeddings (available from index metadata), and computes user--item cosine similarity features in both embedding spaces (ID-based and text-based). These personalization features are combined with lexical and behavioral signals in the final ranking model. The end-to-end personalization adds minimal latency to the non-personalized baseline.

%% ==================================================================
\section{Embedding Models}
\label{sec:embeddings}
%% ==================================================================

\subsection{Problem Formulation}

Let $\mathcal{V}$ denote the set of all video items (movies and TV shows), and let $\phi(v) \in \mathbb{R}^d$ be the embedding of item $v \in \mathcal{V}$. Training instances are constructed from search-session sequences. For a current item $i_t$, we define a positive item $i_{t+1}^{+}$ as the next watched/converted item from the same user within a specified period of time, and a negative item $i^{-}$ from in-batch negative sampling. The objective is to make the positive pair more similar than the negative pair by a margin $m>0$, implemented via the pairwise hinge loss
\begin{equation}
\mathcal{L}=\max\!\Big(0,\, m - \cosine\!\left(\phi(i_t), \phi(i_{t+1}^{+})\right) + \cosine\!\left(\phi(i_t), \phi(i^{-})\right)\Big),
\label{eq:obj}
\end{equation}
which serves as the canonical training objective for both embedding models.

\subsection{Training Data Generation}
\label{sec:emb_data}

Training data for both embedding models is derived from \emph{search logs}, a dataset of individual user search interactions recording queries, impressed items, clicked items, and converted (purchased/played/subscribed) items alongside temporal metadata.

\textbf{Positive pairs.} A positive (anchor, positive) pair is formed when two items, $v_a$ and $v_p$, appear in the same user's search-driven engagement history within a specified time window. We do not use general playback-log data for this step. Although the broader playback corpus is substantially larger, it is noisier (e.g., it includes recommendation and homepage exposures), whereas search-log-derived pairs are smaller but better aligned with the search use case.

\textbf{Negative sampling.} For each positive pair, negative items are sampled uniformly from the current mini-batch.

\textbf{LLM-based evaluation data.} Session-derived evaluation is biased by the same factors that confound training: shared-account co-engagement spans divergent genres, promoted and free-tier content drives co-viewing independent of similarity, and head items dominate the co-engagement graph. We construct a held-out evaluation corpus using a large instruction-tuned language model (LLM) to decouple embedding quality assessment from these exposure biases. Anchor items are sampled uniformly across popularity deciles, and each item appears as an anchor a bounded number of times to ensure balanced representation across the catalog---a property unattainable with session-derived pairs. The LLM is prompted with a role specification, a one-shot contrastive example, and structured metadata (title, media type, genre, principal cast) for each candidate pair, and returns a binary similarity judgment under the criterion: ``two items are similar if a viewer of one is likely to enjoy the other.'' Positive judgments form the positive set for triplet evaluation; negatives form high-quality hard negatives. The corpus is used exclusively for evaluation; no LLM call is made at serving or training time.

\subsection{Text-Based Embedding Model: \textemb{}}
\label{sec:elise}

\begin{figure}[ht]
\centering
\begin{tikzpicture}[
  node distance=0.45cm and 0.8cm,
  inp/.style={rectangle,draw,fill=gray!15,minimum width=3.3cm,minimum height=0.55cm,font=\small,align=center},
  enc/.style={rectangle,draw,fill=blue!15,minimum width=3.3cm,minimum height=0.6cm,font=\small,align=center},
  proj/.style={rectangle,draw,fill=green!15,minimum width=3.3cm,minimum height=0.55cm,font=\small,align=center},
  arr/.style={->,thick}
]
\node[inp] (fields) {Title, Genre,\\Cast/Crew, Synopsis etc.};
\node[enc, above=of fields] (tokenize) {Field-wise tokenization\\(fixed length)};
\node[enc, above=of tokenize] (encoder) {Shared text encoder\\(applied per field)};
\node[enc, above=of encoder] (fieldemb) {Reshape/Concatenated embeddings\\$k \times d_{\mathrm{enc}} \rightarrow k\, d_{\mathrm{enc}}$};
\node[proj, above=of fieldemb] (mlp) {BN $\rightarrow$ Linear $\rightarrow$ ReLU\\$\rightarrow$ BN $\rightarrow$ Linear};
\node[proj, above=of mlp] (out) {L2-normalized item embedding\\$\phi_t(v) \in \mathbb{R}^{d}$};
\draw[arr] (fields) -- (tokenize);
\draw[arr] (tokenize) -- (encoder);
\draw[arr] (encoder) -- (fieldemb);
\draw[arr] (fieldemb) -- (mlp);
\draw[arr] (mlp) -- (out);
\end{tikzpicture}
\caption{\textemb{} text-based item embedding architecture. Item metadata fields are encoded with a shared text encoder, transformed, passed through an MLP head, and L2-normalized for cosine scoring.}
\Description{Text embedding model pipeline showing metadata fields, field-wise tokenization, a shared text encoder applied per field, concatenated embeddings, an MLP head, and an L2-normalized item embedding.}
\label{fig:textemb}
\end{figure}
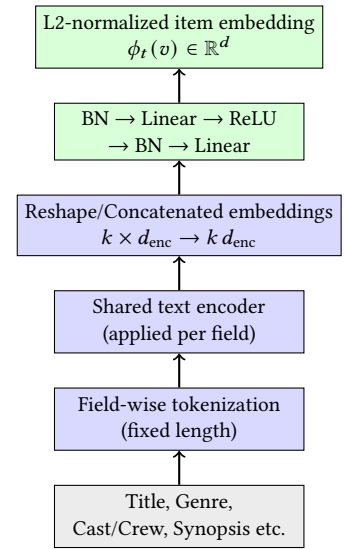

The text-based embedding model (\textemb{}) is a Siamese encoder trained on co-played item pairs. For each item, metadata fields are encoded with a shared text encoder and then combined through reshape/concatenation plus an MLP projection head to produce a shared embedding. Figure~\ref{fig:textemb} illustrates the metadata encoding and projection pipeline used to produce the final item vector.

\textbf{Backbone.} The text encoder is a multilingual transformer-based sentence encoder initialized from a strong multilingual backbone and further adapted for semantic similarity over multilingual text pairs. Using a multilingual encoder naturally handles non-English locales without requiring separate models per language.

\textbf{Item Tower.} For each item, we tokenize multiple text fields (genre, title, artists, entity type) independently at fixed length. Each field is encoded by the shared multilingual transformer backbone. The resulting field embeddings are reshaped/concatenated and passed through a projection head (batch normalization, linear layers, and ReLU), then L2-normalized to produce the final item vector. During fine-tuning, the encoder remains trainable so the representation adapts to domain-specific semantics in media search.

\textbf{Training objective.} Both embedding models optimize the pairwise hinge loss in Eq.~\ref{eq:obj}. In implementation, we compute similarity with a dot product between L2-normalized item vectors; under this normalization, dot product is equivalent to cosine similarity, so Eq.~\ref{eq:obj} and the training-time implementation are identical. Here $i_t$ denotes the anchor item, $i_{t+1}^{+}$ the positive item, $i^{-}$ the negative item, $\phi(\cdot)$ the embedding function, and $m$ the margin selected on a validation split.

\textbf{Training Setup.} The model is trained for multiple epochs using multi-GPU mixed-precision training, with tuned batch size and learning-rate schedule selected on a validation split. 

\subsection{ID-Based Collaborative Embedding Model: \idem{}}
\label{sec:idemb}

\begin{figure}[ht]
\centering
\begin{tikzpicture}[
  node distance=0.4cm and 0.8cm,
  enc/.style={rectangle,draw,fill=purple!15,minimum width=3cm,minimum height=0.6cm,font=\small,align=center},
  proj/.style={rectangle,draw,fill=purple!8,minimum width=3cm,minimum height=0.5cm,font=\small,align=center},
  inp/.style={rectangle,draw,fill=gray!15,minimum width=3cm,minimum height=0.5cm,font=\small,align=center},
  arr/.style={->,thick}
]
\node[inp]  (id)   {Item Identifier (Id)};
\node[enc, above=of id] (lookup) {ID Embedding Lookup\\(trainable $|V|\times d_{\text{high}}$)};
\node[enc, above=of lookup] (mlp)   {MLP: $d_{\text{high}} \to d_{\text{mid}} \to d_{\text{mid}} \to d$};
\node[proj, above=of mlp] (iemb)  {ID Embedding $\phi_{id}(v) \in \mathbb{R}^{d}$};
\draw[arr] (id) -- (lookup);
\draw[arr] (lookup) -- (mlp);
\draw[arr] (mlp) -- (iemb);
\end{tikzpicture}
\caption{\idem{} ID-based collaborative embedding architecture. A high-dimensional ID lookup is projected into the shared embedding space, matching \textemb{} for unified cosine scoring.}
\Description{Collaborative embedding training flow using item IDs, embedding lookup, positive and negative pairs, and pairwise hinge-loss optimization.}
\label{fig:idemb}
\end{figure}
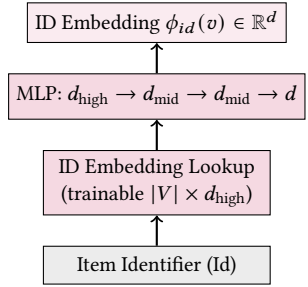

The ID-based model (\idem{}) is also a Siamese encoder trained on co-played item pairs, using only discrete item identifiers (no content metadata). This design captures behavioral co-play patterns that may not be reflected in textual signals e.g. two genre-distinct films that attract the same audience demographic.

\textbf{Architecture.} A trainable ID-embedding lookup maps each item identifier to a high-dimensional latent vector, which is then refined by a compact multilayer projection network into the shared embedding space used for scoring. This design preserves item-level behavioral diversity while producing representations compatible with downstream ranking features.

\textbf{Training.} The same co-played-item pair construction, Siamese training objective, and infrastructure as \textemb{} are used, but the encoder tower replaces text encoding with ID embedding lookup. The model is trained for multiple passes over the data, and optimization settings are aligned with those used for \textemb{} to enable consistent comparison.

\textbf{Coverage limitation and cold-start fallback.} Because the ID table is learned from historical interactions, items not observed during \idem{} training do not receive an ID embedding at all, and very low-frequency items may receive less reliable representations. For such items, the system falls back to the \textemb{} text embedding features at serving time; coverage is summarized in Table~\ref{tab:embedding_coverage}.

\begin{table}[ht]
\centering
\caption{Embedding availability coverage in training data. Session coverage is the fraction of training sessions with at least one item having an available embedding; user coverage is the fraction of users with at least one such item in history.}
\label{tab:embedding_coverage}
\begin{tabular}{lcc}
\toprule
\textbf{Embedding Type} & \textbf{Session Coverage} & \textbf{User Coverage} \\
\midrule
\textemb{} & 100.00\% & 100.00\% \\
\idem{}   & 87.70\% & 86.65\% \\
\bottomrule
\end{tabular}
\end{table}

Table~\ref{tab:embedding_coverage} motivates the hybrid design: \textemb{} provides full catalog/user coverage, while \idem{} contributes high-precision collaborative signal for the majority of sessions and users where interaction-derived items are available. The missing 12--13\% of session/user coverage for \idem{} comes primarily from items that were not seen during item-ID model training, and therefore have no learned ID vector at serving time; this is exactly the gap that \textemb{} covers.

\subsection{Embedding Evaluation}
\label{sec:emb_eval}

Table~\ref{tab:emb_loss} first reports the training and validation losses tracked during model training for both models.

\begin{table}[ht]
\centering
\caption{Embedding-model loss metrics. Lower loss is better.}
\label{tab:emb_loss}
\begin{tabular}{lcc}
\toprule
\textbf{Model} & \textbf{Train Loss} & \textbf{Eval Loss} \\
\midrule
\idem{}         & 0.135 & 0.116 \\
\textemb{}      & 0.136 & 0.124 \\
\bottomrule
\end{tabular}
\end{table}

Table~\ref{tab:emb_eval} summarizes triplet evaluation metrics for both models on the held-out LLM-annotated evaluation corpus.

\begin{table}[ht]
\centering
\caption{Embedding model evaluation on held-out LLM-annotated triplets. Margin = Avg.\ A-P Sim.\ $-$ Avg.\ A-N Sim. Higher triplet accuracy and margin, lower eval loss are better.}
\label{tab:emb_eval}
\begin{tabular}{lcccc}
\toprule
\textbf{Model} & \textbf{Triplet Acc.} & \textbf{Avg A-P} & \textbf{Avg A-N} & \textbf{Margin} \\
\midrule
\idem{}         & 86.02\% & 0.5992 & 0.2295 & 0.3697 \\
\textemb{}      & 89.50\% & 0.6206 & 0.2082 & 0.4124 \\
\bottomrule
\end{tabular}
\end{table}

The results highlight complementary strengths: \idem{} has lower pairwise-loss values, while \textemb{} achieves higher triplet accuracy and a larger A-P/A-N margin on the LLM-labeled set; we defer detailed interpretation to Section~\ref{sec:aggregation_ablation}.

%% ==================================================================
\section{Personalization Ranking Model}
\label{sec:ranker}
%% ==================================================================
\subsection{Feature Engineering}

The personalized ranker extends the existing non-personalized feature set with two embedding-derived affinity features. In practice, features fall into few core groups: long and short-horizon behavioral signals (e.g., click/watch tendencies), lexical text-match signals for query title/cast/genre overlap, catalog-quality signals (e.g., popularity and quality priors), and personalization signals. This preserves the relevance-first behavior of the ranker while adding user-specific evidence.

The two added personalization features encode user--item affinity from collaborative and text-semantic spaces, respectively. They contribute information that is complementary to behavioral aggregates and lexical matching, enabling meaningful per-user reranking even when many candidates have similar global popularity or text evidence, especially for ambiguous and short-prefix queries.

\subsection{Personalization Feature Computation}

At serving time, the affinity score between a user and an item is computed as
\begin{align}
  s_t(u, v) &= \cosine\!\left(\bar{\phi}_t(u),\; \phi_t(v)\right) \\
  s_{id}(u, v) &= \cosine\!\left(\bar{\phi}_{id}(u),\; \phi_{id}(v)\right),
\end{align}
where $\phi_t$ and $\phi_{id}$ are the text and ID embeddings, respectively, and
\begin{equation}
  \bar{\phi}(u) = \frac{1}{|\mathcal{H}_u|}\sum_{v_i \in \mathcal{H}_u} \phi(v_i),
  \label{eq:mean_pool}
\end{equation}
is the mean-pooled user embedding over watch history $\mathcal{H}_u$.

\textbf{Aggregation strategy.} Mean pooling (Eq.~\ref{eq:mean_pool}) was chosen as a deliberate baseline because it is simple, compatible with strict serving-latency requirements, and consistent with a user-representation strategy used in prior large-scale industrial recommendation systems~\cite{covington2016deep}. These properties make mean pooling a natural starting point before introducing more complex sequential user models.

\textbf{Fallback.} When a user has no history or has opted out of personalization, $s_t$ and $s_{id}$ are set to \emph{``missing,''} effectively signaling the ranker to disregard these features while preserving model structure for all other users.

\subsection{Ranker Architecture and Training}

The ranking model is a gradient-boosted decision tree (GBDT) trained with XGBoost~\cite{chen2016xgboost} using a pairwise ranking objective:
\begin{equation}
  \mathrm{Obj} = \sum_{(i,j) \in \mathcal{P}} \log\!\left(1 + e^{-(s_i - s_j)}\right) + \Omega(F)
\end{equation}
where $\mathcal{P}$ is the set of preference pairs derived from relevance labels, $s_i$ is the score for item $i$, and $\Omega(F)$ is XGBoost's regularization term (L1 + L2 on leaf weights).

\textbf{Training data.} The ranker is trained on a large, temporally split corpus of labeled $(\mathrm{query}, \mathrm{user}, \mathrm{candidate})$ examples spanning millions of interactions and a held-out validation set, which supports stable generalization estimates. Labels are weighted to prioritize high-intent outcomes, i.e., $w_{\mathrm{conv}} > w_{\mathrm{click}} > w_{\mathrm{imp}}$, aligning optimization with business impact rather than raw engagement volume. Negative examples are drawn from items in the recall set that received no engagement. The model uses a rich feature set spanning behavioral, lexical, catalog-quality, and personalization signals, including prefix-query traffic.

%% ==================================================================
\section{Offline Evaluation}
\label{sec:offline}
%% ==================================================================

\subsection{Evaluation Protocol}

Offline evaluation is conducted on a strictly temporally held-out test split that follows the training window, preventing label leakage. The evaluation set comprises large-scale search sessions across major device families.

\textbf{Metrics.} We report:
\begin{itemize}
  \item \textbf{NDCG@10}: Normalized Discounted Cumulative Gain at cut-off 10, where items are ordered by personalized score and ground truth is derived from engagement labels.
  \item \textbf{MRR@10}: Mean Reciprocal Rank of the first converted item.
  \item \textbf{Avg.\ Converted Position (ACP)}: Mean rank of the converted item under the model's scoring.
\end{itemize}

\subsection{Ranker Comparison}

Tables~\ref{tab:offline_loss} and~\ref{tab:offline} present the offline evaluation comparing four conditions: (1) the non-personalized baseline (No P13N), (2) the proposed personalized ranker (Ours: \textemb{}+\idem{}), (3) a text-only ablation (ranker with \textemb{} features only, \idem{} removed), and (4) an ID-only ablation (ranker with \idem{} features only, \textemb{} removed).

\begin{table}[ht]
\centering
\caption{Offline loss comparison across baseline and ablations (lower is better).}
\label{tab:offline_loss}
\begin{tabular}{lcc}
\toprule
\textbf{Model} & \textbf{Train Loss} & \textbf{Test Loss} \\
\midrule
No P13N    & 0.2921 & 0.2920 \\
Only Text  & 0.2705 & 0.2725 \\
Only ID    & 0.2678 & 0.2701 \\
\textbf{Text + ID}  & \textbf{0.2662} & \textbf{0.2700} \\
\bottomrule
\end{tabular}
\end{table}

\begin{table}[ht]
\centering
\caption{Offline ranking comparison across baseline and ablations. Higher is better for NDCG@10 and MRR@10; lower is better for average converted-item position.}
\label{tab:offline}
\begin{tabular}{lccc}
\toprule
\textbf{Model} & \textbf{NDCG@10} & \textbf{MRR@10} & \textbf{Avg. Converted Position} \\
\midrule
No P13N    & 0.7113 & 0.6667  & 4.7306 \\
Only Text  & 0.7307 & 0.6868 & 4.0464 \\
Only ID    & 0.7311 & 0.6873  & 4.0587 \\
\textbf{Text + ID}  & \textbf{0.7326} & \textbf{0.6887}  & \textbf{3.9802} \\
\bottomrule
\end{tabular}
\end{table}

\textbf{Key observations.}
\begin{enumerate}
  \item \emph{Hybrid is best overall.} The Text + ID model achieves the strongest results across ranking quality and converted-item position, outperforming both single-signal ablations.
  \item \emph{Complementarity of text and collaborative signals.} Only Text and Only ID both improve substantially over No P13N, but combining them yields the highest NDCG@10 and MRR@10, indicating complementary signal value.
  \item \emph{Better conversion placement.} Average converted-item position improves from 4.7306 (No P13N) to 3.9802 (Text + ID), moving converted items meaningfully closer to the top of the ranked list.
  \item \emph{Generalization quality.} Train/test losses are lowest for the hybrid model, consistent with stronger ranking quality on held-out data.
\end{enumerate}

\textbf{Aggregate summary.} Relative to No P13N, the hybrid model improves NDCG@10 and MRR@10 while reducing average converted-item position, with gains beyond either single-signal ablation.

\subsection{Embedding Quality Comparison}
\label{sec:aggregation_ablation}

Table~\ref{tab:emb_position} reports standalone retrieval quality: the average rank of the converted item when candidates are ordered by user--item cosine similarity alone, without the downstream ranker. \idem{} marginally outperforms \textemb{} on session data (10.81 vs.\ 11.05 all-recalled; 1.89 vs.\ 1.91 impressed). Table~\ref{tab:emb_eval}, in contrast, shows \textemb{} winning on LLM-judged triplet accuracy (89.50\% vs.\ 86.02\%) and A-P/A-N margin (0.4124 vs.\ 0.3697).

This inversion is not a contradiction but the signature of two signals optimized for different notions of similarity, measured by evaluations with different biases. \textemb{}'s contrastive objective over metadata pushes it toward text-grounded semantic proximity, which closely mirrors what the LLM is asked to judge---the LLM has no visibility into user behavior and evaluates pairs purely from structured metadata. \idem{}, trained only on co-engagement, encodes behavioral proximity: items that attract the same audience regardless of textual overlap. Session-based retrieval naturally favors this behavioral signal because its scoring target (the next converted item) is itself behavioral. Each evaluation is biased toward the embedding whose training objective most closely matches its scoring criterion.

Qualitatively, \idem{} excels on head and mid-tail content where co-engagement is dense (franchise loyalty, cross-genre audience affinity), while \textemb{} generalizes better to long-tail and newly released items where behavioral signal is sparse. The ranker ablation in Table~\ref{tab:offline} confirms these are complementary rather than substitutable: Text+ID outperforms both single-signal variants on every ranker metric, with non-redundant contributions from each feature. We view the LLM corpus and session-based retrieval as two lenses on embedding quality---neither dominates---and the decision to deploy both models is ultimately justified by downstream ranker performance, not by either evaluation alone.

\begin{table}[ht]
\centering
\caption{Average converted-item rank when candidates are ordered by embedding cosine similarity only (lower is better).}
\label{tab:emb_position}
\begin{tabular}{lcc}
\toprule
\textbf{Model} & \textbf{All Recalled Items} & \textbf{Impressed Items Only} \\
\midrule
\textemb{} (text) & 11.05 & 1.91 \\
\idem{} (ID)     & 10.81 & 1.89 \\
\bottomrule
\end{tabular}
\end{table}

\subsection{History Length Sensitivity}
\label{sec:history_sensitivity}

\begin{figure}[ht]
\centering
\includegraphics[width=\columnwidth]{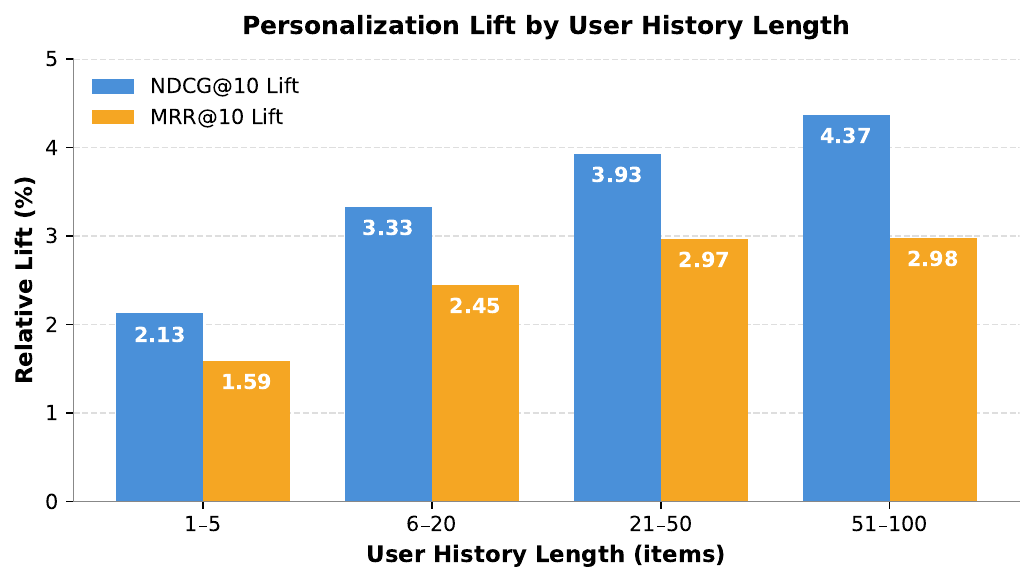}
\caption{Relative lift of the personalized ranker over the non-personalized baseline by user history length. NDCG@10 lift rises monotonically from 2.13\% to 4.37\%; MRR@10 lift also increases with history depth.}
\Description{Bar chart showing relative lift by user history length. NDCG@10 lift increases monotonically from 2.13 percent to 4.37 percent, and MRR@10 lift increases from 1.59 percent to 2.98 percent across four history buckets from 1--5 to 51--100.}
\label{fig:history_length_lift}
\end{figure}

Figure~\ref{fig:history_length_lift} shows relative improvement as a function of user history length, partitioned into buckets. The figure makes the monotonic relationship visually explicit: richer user history strengthens personalization effectiveness.

Gains increase with history depth. Both NDCG and MRR lifts are positive across all buckets and rise from the 1--5 cohort to the 51--100 cohort, indicating that richer watch histories enable stronger personalization while still yielding measurable improvements for sparse-history users. Notably, baseline relevance is lower for longest-history cohorts (NDCG@10 0.733 → 0.680), so the larger lift suggests personalization adds the most value where the default ranker underperforms.

\subsection{Query-Length Sensitivity}
\label{sec:query_length_sensitivity}

Table~\ref{tab:query_length} compares non-personalized and personalized ranking quality on short-prefix queries (1--3 characters) versus longer queries (>3 characters).

\begin{table}[ht]
\centering
\caption{Offline impact of personalization by query length. Relative improvement is computed for Text+ID vs.\ No P13N.}
\label{tab:query_length}
\resizebox{\columnwidth}{!}{%
\begin{tabular}{lcccc}
\toprule
\textbf{Model} & \textbf{NDCG@10} & \textbf{MRR@10} & \textbf{NDCG@10} & \textbf{MRR@10} \\
& \textbf{Prefix (1--3)} & \textbf{Prefix (1--3)} & \textbf{Long (>3)} & \textbf{Long (>3)} \\
\midrule
No P13N   & 0.5515 & 0.4856 & 0.7722 & 0.7357 \\
Text + ID & 0.5991 & 0.5337 & 0.7835 & 0.7477 \\
\midrule
Rel. improvement (\%) & 8.63 & 9.91 & 1.46 & 1.63 \\
\bottomrule
\end{tabular}%
}
\end{table}

The gains are larger on short-prefix queries than on longer queries, consistent with higher ambiguity in the earliest incremental-search inputs.

%% ==================================================================
\section{Online A/B Experiment}
\label{sec:ab}
%% ==================================================================

\subsection{Experiment Setup}

We conducted a three-week online A/B experiment, with treatment and control randomized at the user level.

\begin{table}[ht]
\centering
\caption{A/B experiment configuration.}
\label{tab:ab_setup}
\begin{tabular}{ll}
\toprule
\textbf{Parameter} & \textbf{Value} \\
\midrule
Control       & Non-personalized baseline ranker (XGBoost) \\
Treatment     & Personalized ranker (\idem{} + \textemb{} + XGBoost) \\
Traffic split & Equal traffic split between Control and Treatment \\
Region        & Single region \\
Platforms     & Major client platforms \\
Duration      & Three-week experiment \\
\bottomrule
\end{tabular}
\end{table}

\textbf{Eligibility.} Users are eligible for personalization if they have opted in to personalized recommendations.

\textbf{Statistical testing.} Randomization is performed at the user level. We report relative lift with two-sided significance testing with multiple-testing correction across the experiment's metric suite (Section~\ref{sec:ab_results}) and monitor metric stability daily to guard against transient novelty effects.

\subsection{Results}
\label{sec:ab_results}

\begin{table}[ht]
\centering
\small
\caption{Online A/B experiment results with relative effects and corrected hypothesis tests. Relative lifts are reported with 95\% confidence intervals where applicable.}
\label{tab:ab_results}
\begin{tabular}{p{0.34\linewidth}p{0.34\linewidth}p{0.22\linewidth}}
\toprule
\textbf{Metric} & \textbf{Relative Lift \& 95\% CI} & \textbf{Significance} \\
\midrule
Tap-Through Rate (Primary) & \textbf{+1.14\%} ($\uparrow$); [ +0.95\%, +1.32\% ] & $p < 0.001$; $\alpha = 0.04$ \\
Conversion Rate & \textbf{+1.23\%} ($\uparrow$); [ +0.97\%, +1.50\% ] & $p < 0.001$; $\alpha = 0.0033$ \\
Avg.\ Converted Position & \textbf{$-$2.91\%} ($\downarrow$, better) & Descriptive only \\
\midrule
\multicolumn{3}{p{0.94\linewidth}}{\emph{Note:} Per-metric $\alpha$ thresholds reflect multiple-testing correction across the experiment's full guardrail and primary metric suite.} \\
\bottomrule
\end{tabular}
\end{table}

Both primary online metrics cross their corrected significance thresholds with confidence intervals entirely above zero: Tap-Through Rate improves by +1.14\% and Conversion Rate by +1.23\%. Average Converted Position also improves by $-$2.91\% (lower is better), reported as a descriptive ranking-quality indicator. Because most users in both arms were eligible for personalization, these top-line lifts are averaged over the full user population and likely understate the effect for users with usable watch history.

%% ==================================================================
\section{Discussion and Limitations}
\label{sec:discussion}
%% ==================================================================

\textbf{Incremental search is the main personalization regime.} The concentration of lift on short-prefix queries is the paper's strongest result: Table~\ref{tab:query_length} shows +8.63\% NDCG@10 on 1--3 character queries versus +1.46\% on longer queries, consistent with the idea that early prefixes create a low-lexical-information regime where personalization must supply the missing discrimination.

\textbf{Hybrid embeddings are complementary.} Table~\ref{tab:offline} shows that neither embedding family alone is sufficient: \idem{} is strongest where behavioral signal is dense, while \textemb{} broadens catalog coverage and supports long-tail or newly launched content. The LLM-based evaluation provides a second lens on this complementarity by separating semantic similarity from exposure and click bias.

\textbf{Limitations.} Online results are from a single regional deployment; broader generalization should be validated separately. Offline evaluation remains subject to exposure bias; users with sparse history benefit less than richer-history cohorts; and the current mean-pooled user representation does not model temporal dynamics or session intent. Training/serving skew and embedding freshness may further attenuate measured online gains, and this is likely the most important caveat when interpreting offline-to-online fidelity because serving-time history updates and embedding snapshots are not perfectly synchronized with the materialized offline data.

%% ==================================================================
\section{Conclusion and Future Work}
\label{sec:conclusion}
%% ==================================================================

We presented a personalization framework for incremental video search and argued that this setting is a distinct retrieval-and-ranking regime: at 1--3 characters, lexical evidence is too weak to separate plausible candidates, and user-specific signals become disproportionately valuable. Our results support this framing, with the largest gains occurring exactly in that short-prefix regime, and they show that the strongest solution is hybrid: text embeddings provide semantic coverage and cold-start robustness, while ID embeddings contribute high-precision behavioral affinity where interaction data are rich. Future work includes sequential and query-aware user models, multimodal item embeddings, and explicit exploration mechanisms.

%% ==================================================================
\begin{acks}
The authors thank Lauren Hauser, Fabian Jaskotka, Monil Parikh, Suraj Jain, Igor Ranitovic, Fei Yu and Zheng Yang for their contributions. Generative AI tools were used only for limited language polishing; all technical content, results, and final manuscript decisions were produced and verified by the authors.
\end{acks}

%% ==================================================================
\bibliographystyle{ACM-Reference-Format}
\bibliography{refs}

\end{document}